\documentclass{elsart3}

\usepackage{graphicx}
\usepackage{amssymb}
\usepackage{amsmath,amsfonts,latexsym}
\usepackage{array,tabularx}
\journal{Solid State Communications}

\DeclareMathOperator{\Ry}{\mathcal{R}y}

\newcommand{\vect}[1]{{\mathbf #1}}

\newcommand{\Frac}[2]{\displaystyle\frac{#1}{#2}}

\begin{document}

\begin{frontmatter}



\title{Condensation and Lasing of Microcavity Polaritons: Comparison
       between two Models}


\author[cambridge]{F. M. Marchetti\corauthref{cor}},
\corauth[cor]{Corresponding author.}
\author[cambridge]{M. H. Szymanska},
\author[cambridge]{P. R. Eastham},
\author[cambridge]{B. D. Simons},
\author[cambridge,losalamos]{P. B. Littlewood}

\address[cambridge]{Cavendish Laboratory, University of Cambridge,
                    Madingley Road, Cambridge CB3 0HE, UK}

\address[losalamos]{National High Magnetic Field
                    Laboratory, Pulsed Field Facility, LANL, Los
                    Alamos, NM87545}

\begin{abstract}
  Condensation of microcavity polaritons and the substantial influence
  of pair-breaking disorder and decoherence leading to a laser regime
  has been recently considered using two different models: a model for
  direct two band excitons in a disordered quantum well coupled to
  light and a model where the cavity mode couples instead to a medium
  of localised excitons, represented by two-level oscillators in the
  presence of dephasing processes. Even if complementary from the
  point of view of assumptions, the models share most of the main
  conclusions and show similar phase diagrams. The issue whether
  excitons are propagating or localised seems secondary for the
  polariton condensation and the way in which pair-breaking disorder
  and decoherence processes influence the condensation and drive the
  microcavity into a lasing regime is, within the approximations used
  in each model, generic. The reasons for the similarities between the
  two physical situations are analysed and explained.
\end{abstract}

\end{frontmatter}


\section{Introduction}
\label{sec:intro}
Since exciton-polaritons in semiconductor microcavities posses very
light effective mass, there has been a large interest in recent years
in realising polariton BEC~\cite{deng}. So far, however, experiments
concentrate on a very low density regime, in which a clear polariton
splitting in a normal state is measured. Although the physical picture
of polariton condensation is more complex at higher densities, recent
works suggest that the fermionic structure of excitons does not
prevent condensation and moreover, at higher densities, the condensate
becomes more robust to pair-breaking decoherence and disorder, real
antagonists of condensation \cite{marzena,francesca}. In the absence
of disorder, even at high excitation densities, a strongly coupled
light-matter condensate can be realised and it is only a strong
influence of the environment (all other processes other than dipole
interaction) which leads to a weak light-matter coupling
characteristic of a laser and governed by a Fermi Golden rule. The
laser regime emerges from the polariton condensate at high densities
when pair-breaking disorder or decoherence is large in a way similar
to gapless superconductivity and it arises because these processes
destroy the electronic polarisation leaving the photon component
unchanged.

This conclusion and several other properties of polariton condensate
has been shown to be independent on whether the excitons are
propagating or localised. Two models have been recently considered,
where the condensation of cavity polaritons and the crossover to a
laser regime is not restricted by the assumption on the bosonic nature
of polaritons~\cite{paul,marzena,francesca}, or, in other words, by
the low density limit. In the first one~\cite{francesca}, direct two
band excitons in a disordered quantum well are affected by Coulomb
interaction and are dipole coupled to the cavity light. In the second
one, instead, excitons are localised, e.g. by disorder, and described
as two-level oscillators coupled to light~\cite{paul}.  Here,
decoherence effects can be introduced by coupling the system to
external baths~\cite{marzena}.

The main aim of this paper is to show and explain that similar
conclusions can be drawn from both approaches.  As far as some of the
assumptions are concerned, the two models are in fact complementary
(see the table~\ref{tab:table}). Firstly, while in the first one
dispersion of electrons and holes is included, in the second one only
the presence or absence of localised excitons is taken into
account. As a result, Coulomb interaction and screening at high
excitonic densities is self-consistently included in the band model,
while in the two-level oscillators version only the on-site Coulomb
interaction is taken into account. Secondly, because of technical
restrictions, we limit the analysis of the band model to the high
density regime, while in the two-level systems model all range of
excitation densities is analysed. Recently, a study of the BEC-BCS
crossover in this model has been accomplished~\cite{jonathan}.
Finally, while in one case we study the effect of a pair-breaking
disorder potential, in the other case the role of decoherence is
considered. Though conceptually very different, the two mechanisms
have a similar influence on the condensation of the
electron-hole/photon system in the cavity and both, when large, drive
it towards the lasing regime. Since the principal aim has been the
study of the influence of decoherence and disorder on the
condensation, both models assume thermal equilibrium which correspond
to such a physical situation when thermalization rate is larger than
the pumping and decay rates.
\begin{table}
\begin{center}
\begin{tabularx}{\linewidth}{||X||X||}
\hline\hline 
\textbf{Band Model} & \textbf{Two-level systems}\\ 
\hline
band dispersion & exciton localisation\\ 
\hline 
Coulomb interaction and screening & on-site Coulomb interaction and
phase space filling \\
\hline
dipole coupling to light &  dipole coupling to light\\
\hline
non-pair-breaking disorder &  inhomogeneous broadening of exciton energies\\
\hline 
pair-breaking disorder  & decoherence (coupling to external
baths)\\  
\hline\hline
\end{tabularx}       
\vspace{0.5em}
\caption{\small Principal components of the `band' model and the
        `two-level systems' one.
        \label{tab:table} 
        }
\end{center}
\end{table} 

The paper is organised as follows: The main results relative to each
model are respectively delineated in the next two sections, while a
comparison is described in the concluding one.

\section{Band Model}
\label{sec:bande}
The Hamiltonian for the coupled electron-hole/photon system can be
separated in the following components:
\begin{equation}
  \hat{\mathcal{H}} -\mu \hat{N}_{\text{ex}} =
  \hat{\mathcal{H}}_{\text{ei}} + \hat{\mathcal{H}}_{\text{dis}} + 
  \hat{\mathcal{H}}_{\text{ph}} + \hat{\mathcal{H}}_{\text{int}} \; .
\label{eq:model}
\end{equation}
The Hamiltonian $\hat{\mathcal{H}}_{\text{ei}}$ represents the
interacting Hamiltonian for a direct-gap semiconducting quantum well:
\begin{multline*}
  \hat{\mathcal{H}}_{\text{ei}} = \sum_{\vect{p}}
  \left(\Frac{\vect{p}^2}{2m} - \Frac{\mu - E_g}{2}\right)
  \left(b_{\vect{p}}^\dag b_{\vect{p}} + a_{\vect{p}}
  a_{\vect{p}}^\dag\right) \\ 
  + \Frac{1}{2} \sum_{\vect{q} \ne 0} v(\vect{q})
  \left(\rho_{\vect{q}} \rho_{-\vect{q}} - \sum_{\vect{p}}
  b_{\vect{p}}^\dag b_{\vect{p}} - \sum_{\vect{p}} a_{\vect{p}}
  a_{\vect{p}}^\dag\right) \; ,
\end{multline*}
where the middle of the gap $E_g$ is taken as the energy reference and
where $\rho_{\vect{q}} = \sum_{\vect{p}} (b_{\vect{p} + \vect{q}}^\dag
b_{\vect{p}} - a_{\vect{p}} a_{\vect{p} + \vect{q}}^\dag)$ is the
total electron density operator ($b_{\vect{p}}^\dag$ and
$a_{\vect{p}}^\dag$ create an electron with momentum $\vect{p}$
respectively in the conductance and valence band). Without loss of
generality, we assume electrons and holes to have the same mass
$m$. In the high density regime of electrons and holes $\rho_{el}
a_0^2 \gg 1$ ($\rho_{el}$ denotes the real density of electrons and
holes and $a_0 = 2 \epsilon_0 /e^2 m$ the excitonic Bohr radius),
Coulomb interaction, $v (\vect{q}) = 4 \pi e^2/[\epsilon_0(\vect{q}^2
+ \kappa^2)]$ is screened due to both electrons and holes and, in two
dimensions, the screening length is set by the Bohr radius, $1/\kappa
= a_0$. Hence, in this limit, the Coulomb interaction can be replaced
by a short range contact interaction~\cite{zittartz,comte_nozieres},
where the coupling constant, $g_c$, will be a decreasing function of
the increasing total density of excitations, $\rho_{\text{ex}}
a_0^2$~\eqref{eq:nexci}. Applying a self-consistent Hartree-Fock
treatment, the Coulomb quartic interaction can be decoupled by means
of the excitonic order parameter or polarisation field, $\Sigma
(\vect{r}) = - g_c L^2 \langle \text{g.s.}| b^\dag (\vect{r}) a
(\vect{r}) |\text{g.s.}\rangle$. Note that ordinary Hartree-Fock
pairings, while crucial in the low-density limit, effect only a small
renormalisation of the single-particle energy. Therefore, the system
shows a BCS-like instability around the Fermi surface, $(\mu -
E_g)/2$, and the condensation in the exciton insulator state is
signalled by the opening of a gap equal to $\Sigma
(\vect{r})$~\cite{keldysh_kopaev_kozlov}.
\begin{figure}
\begin{center}
\includegraphics[width=1\linewidth,angle=0]{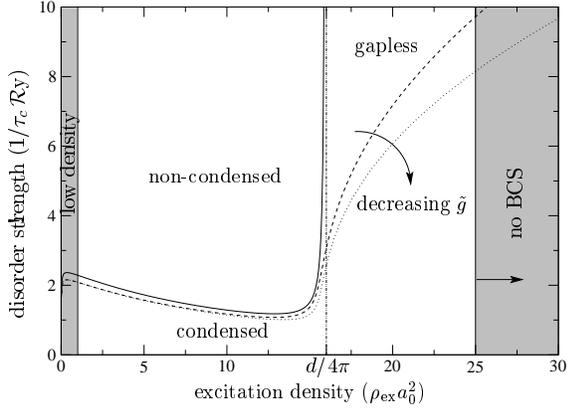}
\end{center}
\caption{\small Zero temperature mean-field phase diagram for the band
         model, for $d=200$ and $\tilde{g} = 1$ (solid and
         dashed). The boundary between the condensed gapped and
         gapless region for $\tilde{g} = 0.8$ (dotted) is shown for
         comparison (in this case the regime where the BCS description
         loses validity starts at $\rho_{\text{ex}} a_0^2 = 31$).}
\label{fig:phase}
\end{figure}
The second term in~\eqref{eq:model} describes the impurity
potentials:
\begin{multline}
  \hat{\mathcal{H}}_{\text{dis}} = \int d\vect{r}\; \left\{\left[U_c
  (\vect{r}) + U_n (\vect{r})\right] b^\dag (\vect{r}) b (\vect{r})
  \right. \\ 
  \left. + \left[U_c (\vect{r}) - U_n (\vect{r})\right]a^\dag
  (\vect{r}) a (\vect{r})\right\} \; .
\label{eq:poten}
\end{multline}
Variations in the width of the quantum well through atomic terracing,
strain fields and so on, induce a `charge-neutral' potential $U_n
(\vect{r})$ that act on electrons and holes with the same sign and, if
weak, does not affect the integrity of the condensate phase. In
contrast, the asymmetric `charged' component $U_c(\vect{r})$ describes
alloy impurities and presents a pair-breaking perturbation which
gradually destroys the excitonic condensate. In both cases, we suppose
that the potentials $U_{n,c}$ are drawn at random from a Gaussian
white-noise distribution, with zero mean and variance, $(2\pi \nu
\tau_{n,c})^{-1}$, proportional to the associated inverse scattering
times, $\tau_{n,c}$, where $\nu = m/2\pi$. The pair-breaking
scattering time can be as well interpreted as the time after which the
relative phase of an excitonic pair is randomised. In the limit of
weak disorder, multiple scattering processes from a single impurity
are suppressed and the system can be treated within the
self-consistent Born approximation -- note that while the
consideration of unitarity limit scatterers would change the results
quantitatively, the qualitative conclusions of the present analysis
remain valid. Here, the effect of pair-breaking disorder on the
integrity of the exciton insulator phase (i.e. in absence of photon
interaction) has been explored in an early work by
Zittartz~\cite{zittartz}.

The Hamiltonians describing free photons in
the cavity and the dipole coupling to the electron-hole system are
respectively given by
\begin{align*}
  \hat{\mathcal{H}}_{\text{ph}} &= \sum_{\vect{p}}
  \psi_{\vect{p}}^\dag\left[\omega_{\vect{p}} - \mu\right]
  \psi_{\vect{p}} \\
  \hat{\mathcal{H}}_{\text{int}} &= g \int d\vect{r}\left[\psi
  (\vect{r}) b^\dag (\vect{r}) a (\vect{r}) + \text{h.c.}\right] \; ,
\end{align*}
where the dispersion $\omega_{\vect{p}} = \sqrt{\omega_c^2 +
(c\vect{p})^2}$ is quantised in the direction perpendicular to the
plane of the cavity mirrors. We notice that the decoupling of Coulomb
interaction through the excitonic order parameter has place in the
same channel as the coupling to photon. Therefore, at mean-field
level, the condensate of the polariton system is characterised by an
order parameter which is a combination of the polarisation and photon
amplitudes, $\Delta = \Sigma + g\psi$. Moreover, one can show the
following constraint has to be verified~\cite{francesca}
\begin{equation}
  g_c (\omega_c - \mu) |\psi| = g |\Sigma|\; ,
\label{eq:llock}
\end{equation}
which leads to the following renormalised pairing interaction, $\Delta
  = - g_{\text{eff}} L^2 \langle \text{g.s.}| b^\dag (\vect{r}) a
  (\vect{r}) |\text{g.s.}\rangle$:
\begin{equation}
  g_{\text{eff}} = g_c + \Frac{g^2}{\omega_c - \mu}\; .
\label{eq:reson}
\end{equation}
As a consequence, in the mean-field approximation and in absence of
disorder, the condensation is described in BCS terms, where the
interplay between electron-hole and photon excitations is contained in
the effective coupling constant, $g_{\text{eff}}$, and in the chemical
potential, $\mu$. Finally, we will suppose that the
electron-hole/photon system is held in quasi-equilibrium by tuning the
chemical potential in~\eqref{eq:model} to fix the total number of
excitations
\begin{equation}
  \hat{N}_{\text{ex}} = \sum_{\vect{p}} \psi^\dag_{\vect{p}}
  \psi_{\vect{p}} + \Frac{1}{2} \sum_{\vect{p}}
  \left(b^\dag_{\vect{p}} b_{\vect{p}} + a_{\vect{p}}
  a^\dag_{\vect{p}}\right)\; ,
\label{eq:nexci}
\end{equation}
where the density of excitations will be indicated with
$\rho_{\text{ex}} = N_{\text{ex}} L^2$. However, how the system
chooses to portion the excitations between the electron-hole and
photon degrees of freedom depends sensitively on the properties of the
condensate.

In particular, in the clean limit, it is easy to see that when the
density $\rho_{\text{ex}} a_0^2$ is low, the majority of the
excitations are invested in electrons and holes and therefore the
density increases linearly with the chemical potential. Here, provided
the density is large enough to keep the particles unbound, the
condensed phase is reminiscent of the exciton insulator one, even if a
small fraction of photons do contribute to the condensate. Therefore,
in this regime the zero temperature phase diagram (see
figure~\ref{fig:phase}) mimics the behaviour of the symmetry broken
exciton insulator: At a fixed value of disorder, the amplitude of the
order parameter decreases increasing the density because of the
Coulomb interaction screening. When the scattering rate $\tau_c$ is
comparable to the value of the unperturbed order parameter, the system
enters a gapless phase before the condensate is extinguished
altogether. Further, adding more excitations into the system,
eventually, as the chemical potential approaches the band edge,
$\omega_c$, the photons are brought into resonance and the character
of the condensate changes abruptly. Screening suppresses the excitonic
coupling constant $g_c$ and at the same time the photon effective
coupling constant $g^2/(\omega_c - \mu)$ grows in size. Here, the
excitations become increasingly photon-like, with $\rho_{\text{ex}}
a_0^2$ diverging exponentially when the chemical potential converges
on $\omega_c$~\cite{francesca}. As a consequence, the condensate
becomes more and more robust against the disorder potential until,
eventually, the complete quenching of coherence is
inhibited. Interestingly, however, the residual effect of the disorder
leaves open the possibility of a substantial region of the phase
diagram where the system is gapless. Here, the condensate manifests
the conventional properties of a semiconductor laser --- i.e. a
substantial coherent optical field, but a gapless spectrum of
electron-hole pairs with negligible electronic polarisation. However,
at sufficiently large densities, the photon-dominated order parameter
becomes comparable with the Fermi energy and the BCS-type description
loses its validity.

Measuring energies in units of the Rydberg ($\Ry =e^2/2\epsilon_0
a_0$), the phase diagram in Fig.~\ref{fig:phase} is characterised by
the total excitation density ($\rho_{\text{ex}} a_0^2$), the disorder
strength ($1/\tau_c \Ry$), the dimensionless photon coupling strength
$\tilde{g} = g (\nu L^2/\Ry)^{1/2}$ and the dimensionless Coulomb
coupling strength, which, fixing the thickness of the quantum well of
the order of the Bohr radius $a_0$, is a function of $x = (\omega_c -
\mu)/\Ry$, in particular $g_c \nu L^2 = 10 [1 + 4
(d-x)]^{-1/2}$. Here, the parameter $d = (\omega_c - E_g)/\Ry$
characterises the crossover between the electron/hole and the photon
dominated region, which is approximatively given by $d/4\pi$. This is
the density of electronic excitations, which, in absence of photons,
can be reached when the chemical potential is equal to $\omega_c$.  In
contrast, fixing the value of $d$, variations in the coupling
$\tilde{g}$ bring small changes in the electron/hole dominated region,
while, for example decreasing its value as shown in
figure~\ref{fig:phase}, it brings to a strong enhancement of the
gapless region. Note as well that diminishing the value of $\tilde{g}$
pushes the strong coupling region to higher densities.
\section{Excitons as Two-Level Oscillators}
\label{sec:2leve}
When the charged component of the disorder potential is negligible,
while the neutral component is strong enough to localise the excitons,
a simple model describing $N$ localised excitons can be rewritten as
\begin{multline}
  \hat{\mathcal{H}}_{\text{ei}} + \hat{\mathcal{H}}_{\text{dis}} +
  \hat{\mathcal{H}}_{\text{int}} = \sum_{j=1}^{N} \left(\epsilon_j -
  \Frac{\mu}{2}\right) \left(b_j^\dag b_j + a_j a_j^\dag\right)\\ 
  + g \sum_{j=1}^{N} \sum_{\vect{p}} \left(e^{- 2\pi i \vect{p} \cdot
  \vect{r}_i} \psi_{\vect{p}} b^\dag_j a_j + \text{h.c}\right)\; ,
\label{eq:twole}
\end{multline}
while the coupling to photons, $\hat{\mathcal{H}}_{\text{ph}}$,
remains unchanged. As before, the system is described by two order
parameters, the coherent photon field and the polarisation, with the
difference that, since Coulomb interaction is contained in the on-site
energies $\epsilon_j$, the polarisation does not have to be
self-consistently determined, but it is locked to the photon field by
the relation~\eqref{eq:llock}. Again the values of the order
parameters depend on the density of excitations, $\rho_{\text{ex}} =
N_{\text{ex}}/N$, or equivalently on the chemical potential, $\mu$. In
addition, one can introduce dephasing processes, coupling the
system~\eqref{eq:twole} to external baths \cite{marzena}:
\begin{multline*}
  \hat{\mathcal{H}}_{\text{bath}} = \sum_{j,\vect{p}}
  \Gamma^{(1)}_{\vect{p}} \left(b^\dag_j b_j - a^\dag_j a_j\right)
  \left(c_{1,\vect{p}}^\dag + c_{1,\vect{p}}\right) \\
  + \sum_{j,\vect{p}} \Gamma^{(2)}_{\vect{p}} \left(b^\dag_j b_j +
  a^\dag_j a_j\right) \left(c_{2,\vect{p}}^\dag +
  c_{2,\vect{p}}\right)\\ 
  + \sum_{\vect{p}} \Omega^{(1)}_{\vect{p}} c_{1,\vect{p}}^\dag
  c_{1,\vect{p}} + \sum_{\vect{p}} \Omega^{(2)}_{\vect{p}}
  c_{2,\vect{p}}^\dag c_{2,\vect{p}}\; .
\end{multline*}
Here, our interest is restricted to processes which do not change the
total number of excitations in the cavity, $\hat{N}_{\text{ex}} =
\sum_{\vect{p}} \psi^\dag_{\vect{p}} \psi_{\vect{p}} + \frac{1}{2}
\sum_{j}(b^\dag_{j} b_{j} + a_{j} a^\dag_{j})$, as, for example,
collisions with phonons. Analogously to disorder in the previous
model~\eqref{eq:poten}, decoherence processes can be divided into
non-pair-breaking ($\Gamma^{(1)}_{\vect{p}}$) and pair-breaking
($\Gamma^{(2)}_{\vect{p}}$) ones.
\begin{figure}
\begin{center}
\includegraphics[width=1\linewidth,angle=0]{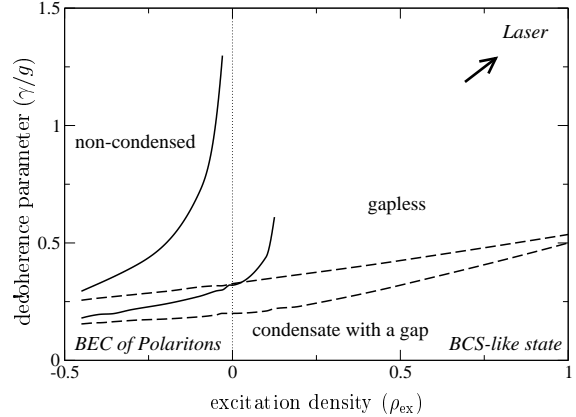}
\end{center}
\caption{\small Zero temperature mean-field phase diagram for the
        localised excitons model. The dashed lines show the phase
        boundaries between the gapped and the gapless condensate,
        while the solid line refer to the boundaries between the
        gapless condensate and the non-condensed system. The upper
        curves correspond to the resonance where
        $\omega_c-2\epsilon=0$, while the lower curves are for
        detuning $\omega_c-2\epsilon=g$, where the photon field is
        above the exciton line by an energy equal to the dipole
        coupling.}
\label{fig:phas1}
\end{figure}
The degrees of freedom of the bath, $c_{1,\vect{p}}$ and
$c_{2,\vect{p}}$, can be integrated out, inducing a quartic
interaction between different two-level systems, which, similarly to
how disorder has been treated in the previous model, can be analysed
within a self-consistent Born approximation. The main difference lies
in the fact that disorder is static, while the baths are characterised
by a dynamics. However, for simplicity, one can assume a static limit,
in which the bath is characterised by only one frequency, e.g. the
lowest one. In this case the decoherence parameter for, say, the
pair-breaking term, will be indicated with $2 \gamma^2 =
\sum_{\vect{k}} \Gamma^2_{\vect{p}}/\Omega_{\vect{p}}$. In the
opposite limit instead, one can assume each mode of the bath
oscillating independently from the other ones. In this limit, called
Markovian approximation, as for the previous one, the mean-field
equations admit an analytical solution. Here we will refer to the
first case.

In absence of decoherence, the minimum $\rho_{\text{ex}} = -0.5$
corresponds to the case where there are no photons and no electronic
excitations. At low excitation densities, where $\rho_{\text{ex}}
\gtrsim -0.5$, the usual bosonic polaritons limit is preserved, while
away from the minimum the condensate gradually becomes more
photon-like, the phase space filling effect of the fermionic states
start to become relevant and the bosonic picture breaks
down~\cite{paul}. In the absence of dephasing, however, even at high
excitation densities, the ground state of the system is a strongly
coupled light-matter condensate with a large gap in the excitation
spectrum, i.e. the incoherent luminescence, very different from a
laser. The influence of the pair-breaking decoherence on the
condensate in this model~\cite{marzena} is similar to the influence of
the pair-breaking disorder in the band model described in the previous
section. This is manifested in the phase diagram in
Fig.~\ref{fig:phas1}. The dephasing processes lead to a suppression of
the electronic polarisation and the gap in the excitation spectrum. At
very low excitation densities and low decoherence there is a gapped
condensate with photon and exciton components comparable in size. As
the decoherence is increased, the system reaches a narrow gapless
region and finally there is a complete suppression of the coherent
fields. As the excitation density is increased, the gapless region
becomes larger. For sufficiently large excitations, the coherent
fields are present even in the presence of very large dephasing. In
this regime the condensate is almost entirely photon-like with very
small excitonic polarisation and the spectrum is gapless, reminiscent
of the laser. In contrast, non-pair-breaking processes which give rise
to inhomogeneous broadening of exciton energies have much weaker, only
qualitative, influence on the coherent fields and do not cause any
transitions.

\section{Discussion}
\label{sec:discu}
A comparison between the models described in the previous two sections
can now be drawn. Despite the fact the two approaches start from
different assumptions on the nature of the excitons, they share the
main results and in particular exhibit a very similar phase
diagram. First of all, the description of the condensate in both cases
is done in terms of an order parameter to which both excitonic and
photonic excitations contribute. In the first model the polarisation
field derives from the self-consistent decoupling of Coulomb
interaction, and, at mean-field level, is locked to the photon field
by the relation~\eqref{eq:llock}, which similarly has place for the
second model. Secondly, in both cases, increasing the density of
excitations, the electron-hole dominated region precedes the one
dominated by photons. While, due to a combination of screening of
Coulomb interaction and a resonance mechanism when the chemical
potential reaches the cavity frequency $\omega_c$ in the first case,
in the second one this is due to the space filling effect of the
fermionic degrees of freedom. Either way, this circumstance causes the
pair-breaking disorder from one side and the pair-breaking (static)
decoherence processes from the other one to act very similarly on the
condensate: Complete quenching of the condensate and therefore loss of
coherence in the system by pair-breaking mechanisms is possible only
in the electron-hole dominated region, while, above a certain density
threshold, coherence is preserved and strong pair-breaking effects
drive the system into a gapless regime, where there is no coherence in
the excitonic component of the order parameter but only in the
photonic one. Here, the system exhibits the characteristic of a
conventional laser. Note that, because of screening, in the band model
the condensed region in the electron-hole dominated region is reduced
by the increase of the density, while this effect is absent in the
two-level systems' model.

There are several important issues not yet considered here.  How would
the phase diagram look like in the presence of realistic band-limited
decoherence characteristic, for example, for acoustic phonons?
Moreover, beyond the mean-field approximation, while the effect of
fluctuations and the spectrum of collective excitations has been
analysed within the band model~\cite{francesca} and in the two level
systems one without decoherence~\cite{jonathan}, a similar study in
presence of decoherence will be the subject of further investigations.

From the analysis above we can therefore conclude that, regardless
whether excitons are propagating or localised, high densities do not
preclude polariton condensation.  On the contrary, since the gap in
the density of state is proportional to the coherent field amplitude,
the condensate becomes more robust at high densities. Real antagonists
of condensation are pair-breaking disorder and decoherence processes.



\paragraph*{Acknowledgements}
  FMM acknowledges the financial support of EPSRC (GR/R95951), MHS the
  financial support of Gonville and Caius College and PRE that of
  Sidney Sussex College. The NHMFL is supported by the National
  Science Foundation, the state of Florida and the US Department of
  Energy.

\end{document}